\def\Z_+={{\bf Z}_+}
\def\Z{{\bf Z}}
\def\C{{\bf C}}

\def\e{\eqno}
\def\l{\ldots}
\def\n{\noindent}
\def\q{\quad}
\def\w{\omega}
\def\H{{\cal H}}
\def\Ht{\tilde{\H}}
\def\la{\lambda}
\def\La{\Lambda}
\def\ra{\rangle}
\def\Ft{\tilde{F}}
\def\Kt{\tilde{K}}

\baselineskip=16pt

\font\twelvebf=cmbx12

\hfill {\bf Preprint JINR E17-10550 (1977)}

\vskip 15mm

{\twelvebf \centerline{LIE ALGEBRAICAL ASPECTS OF THE QUANTUM STATISTICS.}

\centerline{UNITARY QUANTIZATION (A-QUANTIZATION)}}

\vskip 32pt
\noindent
T. D. Palev\footnote{$^{a)}$}{E-mail: tpalev@inrne.acad.bg}

\noindent
Institute for Nuclear Research and Nuclear Energy, 1784 Sofia,
Bulgaria

\vskip 48pt
\noindent
It is shown that the second quantization axioms can, in
principle, be satisfied with creation and annihilation
operators generating (in the case of $n$ pairs of such operators)
the Lie algebra $A_n$ of the group $SL(n+1)$. A concept of the
Fock space is introduced. The matrix elements of these operators
are found.

\vskip 48pt

\n
{\bf 0. Foreword}

\bigskip
This manuscript was published as a JINR Preprint E17-10550 in
1977. It was accepted for publication in {\it Comm. Math. Phys.}
under the condition to be shortened. Since I never did this, it
remained unpublished. In view of the recent interest in various
new kinds of statistics it seems to me the results bellow may be
still of some interest. I publish them without any changes,
although one could have added now a lot of new references. I
apologize also for the somewhat old fashioned exposition.

\vskip 24pt

\noindent
{\bf 1. Introduction}

\bigskip
In the present paper we study some of the possible
generalizations of the quantum statistics and more precisely of
the second quantization procedure from a Lie algebraical point of
view. The consideration is made in the frame of the Lagrangian
field theory, however the results can be easily extended to other
cases, e.g., to nuclear or solid state physics.

As is known [1], the ordinary quantum statistics can be
considerably generalized if one quantizes the fields according to
a weaker system of axioms, abandoning the usually accepted $C$-number
postulate, i.e., the requirement for the commutator or the
anticommutator of two fields to be a $C$-number. In this case the
anticommutation relations between the Fermi creation and
annihilation operators $f_i^+$ and $f_i^-$
\footnote{$^{b)}$}{Throughout the paper the indices
$\xi, \eta, \epsilon, \delta $ take values $\pm$ or $\pm 1$,
$\{x,y\}=xy+yx$, $[x,y]=xy-yx$.}
$$
\{f_i^\xi, f_j^\eta\}={1\over 4}(\xi-\eta)^2\delta_{ij} \eqno(1)
$$
can be replaced by a weaker system of double commutation relations
for the so-called para-Fermi operators $b_i^\pm$, namely
$$
[[ b_i^\xi,b_j^\eta], b_k^\epsilon]=
{1\over 2}(\eta-\epsilon)^2\delta_{jk}b_i^\xi-
{1\over 2}(\xi-\epsilon)^2\delta_{ik}b_j^\eta. \eqno(2)
$$

The commutation relations (2) exhibit some remarkable Lie algebraical
properties. It turns out that the para-Fermi operators generate
the algebra of the orthogonal group [2]. To make the statement
more precise, consider a finite number of operators
$b_1^\pm,\ldots,b_n^\pm $. Then the linear envelope over $\C$
of the operators
$$
b_i^\xi, \; [b_j^\eta,b_k^\epsilon], \quad i,j,k=1,\l,n \e(3)
$$
is isomorphic to the classical Lie algebra $B_n$ of the
orthogonal group $SO(2n+1)$ [3].

There exists one-to-one correspondence between the
representations of $B_n$ and the representations of $n$ pairs of
para-Fermi operators [4]. Therefore the para-Fermi quantization is
actually a quantization according to representations of the
algebra of the orthogonal group in odd dimension and therefore
may be called an odd-orthogonal quantization. This is an
important point, a first hint that the group theory can in
principle be relevant for the quantum statistics.

The algebras $B_n, \; n=1,2,\l$ constitute one of the four
infinite series of the so-called classical Lie algebras. In the
Cartan notation (which we follow) they are denoted as
$A_n,\; B_n,\;C_n$ and $D_n$ for algebras of rank $n$,
$n=1,2,\l$. The corresponding groups $SL(n),\; SO(2n+1), \;
Sp(2n)$ and $SO(2n)$ are well known and therefore we do not
define them here.

Once the Lie algebraic structure of the para-Fermi statistics is
established, it is natural to ask whether one can quantize
according to representations of the other classical Lie algebras.
In the present paper we consider this question in connection with
the algebra of the unimodular group.

In Sect. 3 we determine the concept of $A-$statistics, i.e.,
statistics with creation and annihilation operators
($a-$operators) that generate the algebra of the unimodular
group. Next (Sect. 4) we define the Fock spaces $W_p,\;p=1,2,\l$
and the selection rules for the $A$-statistics. The integer $p$,
called the order of the statistics, has well defined physical
meaning: this is the maximal number of particles that can exist
simultaneously (lemma 4). Tn Sect. 5 we calculate the matrix
elements of the $a-$operators. In the limit $p \rightarrow
\infty$ the $a-$operators reduce (up to a constant) to Bose
operators.

The mathematics used in the paper is mainly of Lie algebraical
nature. In order to introduce the notation and to make the
exposition reasonably self-consistent, we collect in the next
section some definitions and properties from the Lie algebra theory.

\vskip 24pt

\n
{\bf 2. Preliminaries and notations}

\bigskip
Let $A$ be a semi-simple complex Lie algebra of rank $n$,
$\cal{H}$ - its Cartan subalgebra. By
$\w_i,\; e_{\w_i},\, i=1,2,\l,p$ we denote the roots and the root
vectors of $A$. The roots $\w_i$ are vectors from the conjugate
space $\H^*$ of $\H$. Sometimes it is convenient to consider them
as vectors from $\H$ using the fact that every linear functional
$\lambda^*\in \H^*$ can be uniquely represented in the form
$$
\la^*(h)=(h,\la), \quad \forall \; h\in \H.  \e(4)
$$
Here $(\;,\;)$ is the Cartan-Killing form on $A$ and $\la \in \H$.
The mapping
$$
\theta:\; \la^* \rightarrow \la\equiv \theta \la^*  \e(5)
$$
of $\H^*$ on $\H$ is one-to-one. From now on we consider the
roots or any other linear functionals either as elements from
$\H^*$ or from $\H$, denoting them in both cases by the same
symbol (i.e., for $\la^*$ we write also $\la$).

With this agreement we can write
$$
[h,e_{\w_i}]=\w_i(h)e_{\w_i}=(h,\w_i)e_{\w_i} \quad \forall h\in
\H.
\e(6)
$$
The Cartan-Killing form defines a scalar product in the space
$\H^r$ which is the real linear envelope of all roots;
$\H=\H^r$+i$\H^r$. Let $h_1,\l,h_n$ be an arbitrary covariant
basis in $\H^r$ (and hence a basis in $\H$). The root $\w_i$
is said to be positive (negative) if its first non-zero
coordinate is positive (negative). The simple roots, i.e., those
positive roots which cannot be represented as a sum of other
positive roots, constitute a basis in $\H$. Any positive
(negative) root is a linear combination of simple roots with
positive (negative) integer coefficients.

Consider an arbitrary finite-dimensional irreducible $A-$module
$W$ (i.e., a space where a finite dimensional irreducible
representation of $A$ is realized. The basis $x_1,\l,x_N$ in
$W$ can always be chosen such that
$$
hx_i=\la_i(h)x_i=(h,\la_i)x_i \quad \forall \; h\in \H,\;
i=1,\l,N. \e(7)
$$
Thus, to every basic vector $x_i\in W$ there corresponds an image
$\la \in \H^*$ or $\H$. The vectors $x_i$ are the weight vectors
and their images - the weights of the $A-$module $W$. The mapping
$\tau:\; x_i \rightarrow \la_i$ is surjective and the number of
the vectors $\tau^{-1}(\la_i)$ is called multiplicity of the
weight $\la_i$. Let $e_\w$ be a root vector and $\la_i$ be the
weight of $x_i$. Then $e_\w x_i$ is either zero or a weight
vector with weight $\w + \la_i$. The $A-$module $W$ contains a
unique (up to multiplicative constant) weight vector $x_\Lambda$
with properties $e_{\w_i} x_\La=0$ for all positive roots $\w_i > 0$.
The weight $\La$ of $x_\La$ is the highest weight of $W$. The
multiplicity of $\La$ is one and $W$ is spanned over all vectors
$$
e_{\w_{i_1}}e_{\w_{i_2}}\l e_{\w_{i_m}}x_\La \quad m=1,2,\l, \e(8)
$$
where $\w_{i_1},\l,\w_{i_m}$ are negative roots. Therefore an
arbitrary weight $\La$ is of the form
$$
\la=\La-\sum_{\w_i>0}k_i\w_i   \e(9)
$$
with $k_i$ positive integers and sum over positive (or only
simple) roots.

Let $\pi_1, \l, \pi_n$ be the simple roots of $A$. Then for an
arbitrary weight $\la$ the $n-$tuple $[\la_1,\la_2,\l,\la_n]$
has integer coordinates defined as
$$
\la_i={2(\la,\pi_i)\over (\pi_i,\pi_i)} \quad i=1,2,\l,n. \e(10)
$$

The $n-$ tuple $[\La_1,\l,\La_n]$ corresponding to the highest
weight $\La$ has non-negative coordinates, and it defines the
irreducible representation of $A$ in $W$ up to equivalence. On the
contrary, to every vector $\La \in \H$, such that  $\La_1,\l,\La_n$
defined from (10) are non-negative integers, there corresponds an
irreducible $A-$module. Thus there exists a one-to-one
correspondence between the irreducible (finite-dimensional)
$A-$modules and the set  $[\La_1,\l,\La_n]$ of non-negative
integers. We call  $[\la_1,\l,\la_n]$ canonical co-ordinates of $\la$.

Define an $F-$basis $f_1, f_2,\l,f_n$ in $\H$
(or in $\H^*$) as follows
$$
f_i={2\over (\pi_i,\pi_i)}\pi_i, \quad i=1,\l,n   \e(11)
$$
and let $K=\{f^1,f^2,\l,f^n\}$ be the corresponding dual basis, i.e.,
$f^i(f_j)=(f^i,f_j)=\delta_{ij}$. For an arbitrary $\la\in \H^*$
we have
$$
\la=\sum_i \la(f_i)f^i=\sum_i {2(\la,\pi_i)\over (\pi_i,\pi_i)}f^i \e(12)
$$
and therefore in the $K-$basis the coordinates of every weight
$\la$ coincide with its canonical co-ordinates.

By means of the $F-$basis one can easily calculate the canonical
co-ordinates of an arbitrary weight $\la$. Indeed
$$
f_ix_\la=\la(f_i)x_\la={2(\la,\pi_i)\over (\pi_i,\pi_i)}x_\la \e(12')
$$
and therefore the $i^{th}$ canonical co-ordinate $\la_i$ of $\la$
is an eigenvalue of $f_i$ on $x_\la$. More generally, if $h_1,\l,
h_n$ is any covariant basis in $\H$, then the covariant
co-ordinates $\la_1,\l,\la_n$ of the weight $\la$, i.e., the
co-ordinates of $\la$ in the dual (or contravariant) basis
$h^1,\l,h^n$ are determined from the relation
$$
h_ix_\la=\la(h_i)x_\la=\la_ix_\la . \e(13)
$$

An important property of the set $\Gamma$ of all weights is its
invariance under the Weyl group $S$, which is a group of
transformations of $\H$.
$S=\{S_{\w_i} \vert \w_i \; - \; roots\; of\; \La \} $ is a finite group,
its elements labelled by the roots $\w_i$ of $A$ are defined as follows:
$$
S_{\w_i}.h=h-{2(h,\w_i)\over (\w_i,\w_i)}\w_i \quad \forall \;
h\in \H. \e(14)
$$
The set $\Gamma$ of all weights is characterized by the following
statement: if $\la \in \Gamma$, then
$$
S_{\w_i}\la=\la + j\w_i\in \Gamma, \quad j\;-\;integer . \e(15)
$$
and $\Gamma$ contains also the weights
$$
\la, \la+\w_i,\la+2\w_i,\l,\la+j\w_i. \e(16)
$$
All weights that can be connected by transformations of the Weyl
group are called equivalent. They have the same multiplicity.
Among the equivalent weights there exists only one weight, the
dominant one, the canonical co-ordinates of which are nonnegative
integers.

\vskip 24pt

\n
{\bf 3. Unitary quantization ($A-$quantization)}

\bigskip
In the case of ordinary statistics the second quantization in the
Lagrangian field theory can be performed in different equivalent
ways. One can start, for instance, from the equal-time
commutation relations. For the generalizations we wish to
consider, it is more convenient to follow the quantization
procedure accepted by Bogoljubov and Shirkov [6].

Apart from the fact that the fields become operators and the
requirement for relativistic invariance, their approach is
essentially based on what we call a main quantization postulate:
the energy-momentum vector $P^m$ and the angular-momentum tensor
$M^{mn}, \; m,n=0,1,2,3$ are expressed in terms of the
operator-fields by the same expressions as in the classical
case.

It follows from this postulate, together with the requirement
that the field transforms according to unitary representations of
the Poincare group and the compatibility of the transformation
properties of the field and the state vectors, that the field
$\Psi(x)$ satisfies the commutation relations
$$
[P^m,\Psi(x)]=-i\partial^m \Psi(x). \e(17)
$$
This relation expresses (in infinitesimal form) the translation
invariance of the theory.

To proceed further, it is convenient to pass to the discrete
notation in the momentum space. Consider the field $\Psi(x)$ with
a mass $m$ locked in a cube with edge $L$. For the eigenvalues
$k_n^m$ of the 4-momentum $P^m, \; m=0,1,2,3$, one obtains
$$
k_n^\alpha ={2\pi\over L}n^\alpha, \quad
k_n^0=\sqrt{m^2 +({2\pi\over L})^2[(n^1)^2 +(n^2)^2+(n^3)^2]},\e(18)
$$
where $n=(n^1,n^2,n^3)$, $\alpha=1,2,3$ and $n^\alpha$ runs over all
non-negative integers. In momentum space the relation (17) reads as
follows
$$
[P^m, a_i^\pm]=\pm k_i^m a_i^\pm , \e(19)
$$
where $a_i^+ \;(a_i^-)$ are the corresponding to $\Psi(x)$
creation and annihilation operators and the index $i$ replace all
discrete indices ($n$, spin, charge, etc.).

The commutation relations between the creation and annihilation
operators are usually derived from the translation invariance law
in momentum space (19). We call it {\it initial quantization equation}
(IQE). To determine the commutation relations one has to specify
one more point. Up to now nothing was said about the creation and
the annihilation operators that enter into $P^m$. In the ordinary
theory it is usually accepted that the dynamical variables are
written in a normal-product form and therefore for the Fermi
fields this gives
$$
P^m=\sum_i k_i^m f_i^+ f_i^-, \e(20)
$$
where $f_i^+ \; (f_i^-)$ are the Fermi creation (annihilation)
operators (1).  One can easily check that the initial
quantization equation (with $a_i^\pm =f_i^\pm$) is compatible
with the anticommutation relations (1). This is, however, not the
case for the para-Fermi operators (2), apart from the case of
their Fermi representation. The para-Fermi statistics cannot be
derived from the normal-product form of the dynamical variables.
In order to fulfil (19) Green chose another ordering of the
operators in $P^m$ and in particular for spinor fields he wrote [1]:
$$
P^m={1\over 2}\sum_i k_i^m [b_i^+,b_i^-].   \e(21)
$$

We see that the ordering of the operators in the 4-momentum is
closely related to the corresponding statistics. It is natural to
expect therefore that any other generalization of the statistics
may require new expressions for $P^m$. In order to get a feeling
as to how one can modify $P^m$, we now proceed to derive the
para-Fermi statistics in such a way that later on it will be
possible to generalize the idea to other case.

We start with the expression (20). In order to use a proper Lie
algeraical language (finite-dimensional Lie algebras), suppose
that the sum in (20) is finite,
$$
P^m={1\over 2}\sum_{i=1}^n k_i^m [b_i^+,b_i^-].   \e(22)
$$
This is only an intermediate step. In the final results we let
$n \rightarrow \infty$.

As we have already mentioned, the set $f_1^\pm,\l,f_n^\pm$ of
Fermi creation and annihilation operators (1) generates on
particular representation (we call it {\it Fermi representation})
of the algebra $B_n$. We put now the question: can the expression
(22) be written in such a form that the initial quantization
equation (19) will hold for the Fermi operators, considered as
generators of $B_n$, i.e., independently of the fact that we are
staying in one particular representation of $B_n$ - the Fermi one.
The Lie algebraical reason why (19) does not hold for the
para-Fermi operators is clear. It is due to the fact that the
4-momentum (22) does not belong to $B_n$ since it contains
products of $b_i^+$ and $b_i^-$, which is not a Lie algebraical
operation.  Therefore the IQR, considered as commutation
relation, is not preserved for other representations of $B_n$.
If however the 4-momentum together with the creation and
annihilation can be embedded in a Lie algebra, so that in the
Fermi case $P^m$ reduces to (22), then the  IQR (19) will hold
for any other representation of this algebra.

For this purpose we rewrite the 4-momentum (22) in the following
identical form
$$
P^m=\sum_{i=1}^n k_i^m
({1\over 2}[f_i^+,f_i^-] + {1\over 2}\{f_i^+,f_i^-\}).   \e(23)
$$
Consider the Lie algebra generated from $f_1^\pm,\l,f_n^\pm$ and
$\{f_i^+,f_i^-\}$. Since $\{f_i^+,f_i^-\}=1$, we obtain the algebra
$B_n\oplus I$, where $I$ is the one-dimensional commutative
center. Now $P^m \in B_n\oplus I $ and therefore the commutation
relation (19) holds for any other representation. In other words,
if we substitute in (23) $f_i^\pm \rightarrow b_i^\pm$ and
$\{f_i^+,f_i^-\} \rightarrow {\bf 1}$, i.e., put
$$
P^m=\sum_{i=1}^n k_i^m
({1\over 2}[b_i^+,b_i^-] + {1\over 2}{\bf 1}),   \e(24)
$$
where {\bf 1} is the generator of the center of $B_n\oplus I$,
then the initial quantization condition (19) will be fulfilled
for any representation of $B_n\oplus I$.

The operator
$$
Q^m=\sum_{i=1}^n k_i^m {1\over 2}{\bf 1}   \e(25)
$$
commutes with all  creation and annihilation operators and all elements
of $B_n\oplus I$. Therefore it is a constant within every
irreducible representation and in the particular case of para-Fermi
statistics the second term in (24) can be omitted. Thus, we
obtain the expression (21) for $P^m$, postulated by Green from
the very beginning.

We shall now apply a similar approach for the algebra $A_n$ of
the unimodular group $SL(n+1)$. The nontrivial part is to find an
analogue of the Fermi operators, i.e., operators $a_i^\pm$ that
generate some representation of $A_n$ and fulfil the initial
quantization equation (19) with 4-momentum written (in this
particular representation) in a normal product form. Then we
shall apply the above procedure to enlarge the class od
admissible representations.

First we recall some properties of $A_n$. We consider $A_n$ as a
subalgebra of the algebra $gl(n+1)$ of the general linear group
$GL(n+1)$. The algebra $gl(n+1)$ may be determined as a linear
envelope of the generators $e_{ij},\; i,j=0,1,\l,n$, that satisfy
the commutation relations
$$
[e_{ij}, e_{kl}]=\delta_{jk}e_{il}-\delta_{li}e_{kj}, \quad
i,j,k,l=0,1,\l,n.  \e(26)
$$
Let $\H$ and $\tilde{\H}$ be the Cartan subalgebras of $A_n$ and
$gl(n+1)$, resp. Denote by $env\{X\}$ the linear envelope of an
arbitrary set $X$. In terms of the $gl(n+1)$ generators we have:
$$
\eqalign{
& gl(n+1)=env\{e_{ij}|i,j=0,1,\l,n\},\cr
& A_n=env\{e_{ii}-e_{jj},\; e_{ij}|i\not= j=0,1,\l,n\},\cr
& \tilde{\H}=\{h_i| h_i=e_{ii},\;i=0,1,\l,n\},  \cr
& \H=\{h_i-h_j| h_i=e_{ii},\;i,j=0,1,\l,n\}.  \cr
}\e(27)
$$

For a covariant basis in $\tilde{\H}$ we choose the vectors
($h_i\equiv e_{ii}$)
$$
h_0,h_1,\l,h_n. \e(28)
$$
The algebra $gl(n+1)$ is not semi-simple. Its Killing form is
degenerate and does not determine a scalar product on
$\tilde{\H^r}$. It is convenient to introduce a metric in
$\tilde{\H}$ with the relation
$$
(h_i,h_j)=2(n+1)\delta_{ij}.\e(29)
$$
Restricted on $\H$ this metric coincides with the Cartan-Killing
form on $A_n$.

From (26) and (29) one obtains
$$
[h,e_{ij}]=(h,h^i-h^j)e_{ij} \quad \forall h\in \H,\;
i\not= j=0,1,\l,n, \e(30)
$$
where $h^0,h^1,\l,h^n$ is the contravariant (i.e., the dual to
$h_0,h_1,\l,h_n$) basis in  $\tilde{\H}$. Hence the generators
$e_{ij},\; i\not= j=0,1,\l,n$ are the root vectors of $A_n$.
The correspondence with their roots is
$$
e_{ij} \rightarrow h^i-h^j, \quad i\not= j=0,1,\l,n. \e(31)
$$

In the basis (28) the generators
$$
e_{ij},\quad i<j \; (i>j), \quad i,j=0,1,\l,n \e(32)
$$
are the positive (negative) root vectors of $A_n$. The simple
roots are
$$
\pi_i=h^{i-1}-h^i, \quad i=1,\l,n. \e(33)
$$
Therefore the $F$-basis (11) in this case reads as
$$
f_i={2\over (\pi_i,\pi_i)}\pi_i=h_{i-1}-h_i, \quad i=1,\l,n   \e(34)
$$
We are now ready to define the analogue of the Fermi operators. Let
$E_{ij}, \; i,j=0,1,\l,n$ be $(n+1)-$square matrix with 1 on the
intersection of  $i-$row and $j-$column and zero elsewhere.
Clearly the mapping
$$
\pi: \; e_{ij} \rightarrow E_{ij}, \quad i,j=0,1,\l,n \e(35)
$$
determines a representation of $gl(n+1)$ and hence its
restriction on $A_n$ gives a representation of $A_n$.

The operators
$$
A_i^+=E_{i0},\;\; A_i^-=E_{0i}, \quad i=1,2,\l,n  \e(36)
$$
generate the algebra $A_n$ (in the above representation) since
$$
[A_i^+,A_j^-]=E_{ij},\quad [A_k^+, A_k^-]=E_{kk}-E_{00},
\quad i\ne j, \quad i,j,k=1,\l,n. \e(37)
$$
Moreover for the commutation relations between
$A_1^\pm,\l,A_n^\pm$ and the operator
$$
P^m =\sum_i k_i^m A_i^+A_i^-  \e(39)
$$
we obtain the right expression:
$$
[P^m,A_i^\pm]=\pm k_i^m A_i^\pm. \e(39)
$$
The operators $A_1^\xi,\l,A_n^\xi$ satisfy the initial
quantization equation and can be considered as creation $(\xi=+)$
and annihilation $(\xi=-)$ operators.

The commutation relation (39) does not hold for other
representations of $A_n$. In order to extend the class of the
admissible representations, we represent the 4-momentum (38) like
in the Fermi case, in the form
$$
P^m =\sum_i k_i^m ([A_i^+,A_i^-]+E_{00}).  \e(40)
$$

Consider now the Lie algebra generated from the operators
$A_1^\pm,\l,A_n^\pm$ and $E_{00}$. One can easily show, it is the
algebra $gl(n+1)=A_n\oplus I$. Since $P^m \in gl(n+1)$, the
initial quantization equation (39) holds for any other
representation of $gl(n+1)$.  Hence we may define representation
independent creation and annihilation operators as follows
$$
a_i^+=e_{i0},\;\; a_i^-=e_{0i}, \quad i=1,2,\l,n.  \e(41)
$$
In this case we have to postulate for $P^m$ the expression
$$
P^m =\sum_i k_i^m ([a_i^+a_i^-]+e_{00}).  \e(42)
$$

The operators $a_i^\pm$ are root vectors of $A_n$. The
correspondence with their roots is
$$
a_i^\pm \; \leftrightarrow \; \mp(h^0-h^i), \q i=1,\l,n, \e(43)
$$
and therefore the creation (annihilation) operators are negative
(positive) root vectors. Since any other root
$h^i-h^j,\; i\ne j=1,\l,n$,
$$
h^i-h^j=(h^0-h^j)-(h^0-h^i)
$$
is a sum of the roots of $a_j^-$ and $a_i^+$, the creation and
the annihilation operators generate the algebra $A_n$.

The commutation relations of $A_n$ can be written in terms of
$a_i^\pm$ only. From (26) we obtain
$$
\eqalign{
& [[a_i^+,a_j^-],a_k^+]=\delta_{kj}a_i^+ + \delta_{ij}a_k^+ ,\cr
& [[a_i^+,a_j^-],a_k^-]=-\delta_{ki}a_j^- - \delta_{ij}a_k^- ,\cr
& [a_i^+,a_j^+]=[a_i^-,a_j^-]=0. \cr
}\e(44)
$$

\n
{\it Definition 1.} The operators $a_i^\pm,\;i=1,2,\l$ satisfying
the commutation relations (44) are called $a-${\it operators} and the
corresponding quantization (statistics) {\it unitary or
$A-$quantization (statistics)}.

We observe that the equal-frequency operators commute with each
other. This property helps a lot in all calculations with the
$a-$operators.

\vskip 24pt

\n
{\bf 4. Fock spaces for the a-operators}

\bigskip
We now proceed to study those representations of the $a-$creation
and annihilation operators that possess the main features of the
Fock space representations in the ordinary quantum mechanics. We
continue to consider a finite set of operators. The extension of
the results to the infinite (including continuum) number of
$a-$operators will be evident.

\smallskip
\n
{\it Definition 2.} Let $a_1^\xi,\l,a_n^\xi$ be $a-$creation
$(\xi=+)$ and annihilation $(\xi=-)$ operators. The $A_n-$module
$W$ is said to be a Fock space of the algebra $A_n$ if it fulfills
the conditions:

\smallskip
{\it 1. Hermiticity condition}
$$
(a_i^+)^*=a_i^-, \q i=1,\l,n. \e(45)
$$
\hskip 12mm Here $^*$ denotes hermitian conjugation operation.

\smallskip
{\it 2. Existence of vacuum}. There exists a vacuum vector
$|0\ra \in W $ such that
$$
a_i^-|0\ra=0, \q i=1,\l,n. \e(46).
$$

\smallskip
{\it 3. Irreducibility}. The representation space $W$ is spanned
over all vectors
$$
a_{i_1}^+a_{i_2}^+\l a_{i_m}^+|0\ra, \q m\in N_0. \e(47)
$$
By $N_0$ we denote the set of all non-negative integers. The Fock
space of $A_n$ is called also $A_n-$module of Fock, Fock module
of the $a-$operators or simply Fock module.

\bigskip
\n
{\it Lemma 1.} The hermiticity condition (45) can be satisfied if
and only if the $A_n-$module $W$ is a direct sum of
finite-dimensional modules.

\smallskip
\n
{\it Proof.} The generators of the compact from $su(n+1)$ of $A_n$
read in terms of the $a-$operators as follows
$$
\eqalign{
& E_{0j}=i(a_j^+ + a_j^-),  \cr
& F_{0j}=a_j^- - a_j^+, \cr
& E_{jk}=i[a_j^+,a_k^-]+i[a_k^+,a_j^-], \cr
& F_{jk}=[a_j^+,a_k^-]-[a_k^+,a_j^-]. \cr
}\e(48)
$$
Evidently the generators are antihermitian if and only if (45) holds.

As is known, the antihermitian representations of the compact
forms of the classical algebras are completely reducible. The
irreducible components are finite-dimensional. This proves the
sufficient part. The necessity follows from the observation that
the metric in any irreducible $su(n)$-module can be introduced so
that the generators are antihermitian.

From the complete reducibility and the irreducibility condition
(definition 1) we conclude.

\bigskip\n
{\it Corollary 1.} The Fock spaces are finite-dimensional
irreducible $A_n-$modules.

\smallskip
In the remaining part of the paper by creation and annihilation
operators we always mean $a-$operators. Moreover we fix the
ordering of the basis in $\tilde{\H}$ to be (28). Then the
creation and the annihilation operators $a_i^+$, $a_i^-$ are
negative and positive root vectors. In this case the operators
$a_1^-,\l,a_n^-$ annihilate the highest weight vector $x_\Lambda$
of the Fock space and hence $x_\Lambda$ is one of the candidates
for a vacuum state.

\bigskip
\n
{\it Lemma 2.} Let $W$ be a Fock space of $A_n$. Up to a
multiplicative constant the vacuum state is unique and coincides
with the highest weight vector $x_\Lambda$ of $W$

\smallskip
\n
{\it Proof.} First suppose the vacuum is a weight vector
$x_\lambda \ne  x_\Lambda$. Then the corresponding weights are
also different, $\lambda \ne  \Lambda$. Moreover
$\Lambda >  \lambda$ (i.e., the vector $\Lambda -  \lambda$ is
positive). The irreducibility condition says there exists a polynomial
$P(a_1^+,\l,a_n^+)$ of the creation operators such that
$$
x_\Lambda=P(a_1^+,\l,a_n^+)x_\lambda. \e(49)
$$
Denote by $\w_i$ the root of $a_i^+$. From (49) we have
$$
\Lambda = \lambda +\sum_{i=1}^n k_i\w_i, \q k_i \in N_0. 
$$
This is, however, impossible, since $\Lambda - \lambda >0$ and
$\sum_{i=1}^n k_i\w_i<0$. We conclude that the vacuum cannot be a
weight vector different from $x_\Lambda$.

More generally, suppose $|0\ra \in W$ is a vacuum state different
from $x_\Lambda$. An arbitrary vector $x \in W$ and in particular
$|0\ra$ can be represented uniquely as a sum of weight vectors
$x_{\lambda_i}$ with different weights $\lambda_i$:
$$
|0\ra=\sum_{j=0}^m x_{\lambda_j}, \q  \lambda_i \ne \lambda_j \q if
\q i\ne j. \e(50)
$$
The vectors $x_{\lambda_0},\l,x_{\lambda_m}$ are linearly independent.
The nonzero of the vectors
$a_i^-x_{\lambda_0},\l,a_i^-x_{\lambda_m}$ are also linearly
independent, since they correspond to different weights. Hence
$$
a_i^-|0\ra=0 \;\;{\rm implies} \;\; a_i^-x_{\lambda_j}=0, \;\;
j=0,1,\l,m. \e(51)
$$
Let for definiteness $\lambda_0>\lambda_1>\l>\lambda_m$.
The vector cannot be a vacuum state if $\lambda_0\ne \Lambda$
since clearly there exists no polynomial $P(a_1^+,\l,a_n^+)$ such
that
$$
x_\Lambda=P(a_1^+,\l,a_n^+)|0\ra. \e(52)
$$
Suppose $|0\ra=x_\Lambda +x_{\lambda_1}+\l+x_{\lambda_m}$. Then
(52) can be satisfied only when there exists a monomial
$(a_1^+)^{l_1}\l(a_n^+)^{l_n}$ with the property
$$
x_{\lambda_1}=(a_1^+)^{l_1}\l(a_n^+)^{l_n}x_\Lambda.
$$
This is, however, impossible since for $l_i\ne 0
\;\;a_i^-x_{\lambda_1}\ne 0$ and this contradicts (51).

In the following theorem we prove one convenient criterion for the
$A_n-$module to be a Fock space.

\bigskip
\n
{\it Theorem 1.} The $A_n-$module $W$ is a Fock space if and only
if it is an irreducible finite-dimensional module such that
$$
a_i^-a_j^+x_\Lambda=0 \q i\ne j=1,\l,n. \e(53)
$$
The highest weight vector $x_\Lambda$ is the vacuum of $W$.

\smallskip
\n
{\it Proof.} Let $W$ be a Fock space. Then it is
finite-dimensional irreducible $A_n-$module (corollary 1) and the vacuum
$|0\ra=x_\Lambda$ (lemma 2). The operator $[a_i^-,a_j^+],\;i\ne j$
is a root vector of $A_n$. Its root $h^j-h^i$ cannot be
represented as linear combination of the roots  $-h^0+h^i$ of the creation
operators $a_1^+,\l,a_n^+$. Hence there exists no polynomial
$P(a_1^+,\l,a_n^+)$ of $a_1^+,\l,a_n^+$ such that
$$
[a_i^-a_j^+]x_\Lambda=P(a_1^+,\l,a_n^+)x_\Lambda \ne 0
$$
Since $a_i^-a_j^+x_\Lambda\in W$ it has to be zero,
$a_i^-a_j^+x_\Lambda=0, \; i\ne j$.

The proof of the sufficient part of the theorem is based on the
Poincare-Birkhoff-Witt theorem [7]: Given a Lie algebra $A$ with
a basis $e_1,\l,e_N$. All ordered monomials
$e_1^{j_1}e_2^{j_2}\l e_N^{j_N}$ constitute a basis in the
universal enveloping algebra $U$ of $A$.

Let in the irreducible finite-dimensional $A_n-$module $W$ the
equality (53) holds. Divide the basis elements of $A_n$ into
three groups
$$
\eqalign{
& I= \{a_i^+,\; [a_j^+,a_k^+] |\;j<k;\;i,j,k=1,\l,n\}
\equiv\{e_{-1},e_{-2},\l,e_{-p}\}, \cr
& II=\{a_i^-,\; [a_j^-,a_k^-],\; [a_r^-,a_s^+]
  |\;j<k;\;r\ne s;\;i,j,k,r,s=1,\l,n\}
\equiv\{e_{1},e_{2},\l,e_{q}\}, \cr
& III=\{\w_k|k=1,\l,n\},
}
$$
where $\w_1,\l,\w_n$ is a basis in the Cartan subalgebra $\H$.
Order the elements within each group in an arbitrary way. From
the irreducibility and the Poincare-Birkhoff-Witt theorem it
follows that $W$ is linearly spanned on all vectors
$$
e_{-1}^{i_1}\l,e_{-p}^{i_p}e_1^{j_1}\l e_q^{j_q}\w_1^{k_1}\l\w_n^{k_n}
x_\Lambda. \e(54)
$$
Since $x_\Lambda$ is an eigenvector of the Cartan subalgebra and
the operators from II annihilate $x_\Lambda$, the vector (54) is
non-zero only if $j_1=j_2=\l =j_n=0$. Hence $W$ is spanned on all
vectors
$$
P(a_1^+,\l,a_n^+)x_\Lambda ,
$$
where $P$ is an arbitrary polynomial of the creation operators.
This proves that $W$ is a Fock space with a vacuum $|0\ra=x_\Lambda$.

Now it remains to determine the irreducible $A_n-$modules
satisfying the condition (53). In order to solve this problem, we
consider first some questions from the representation theory of $A_n$.
As we mentioned, it is convenient to consider $A_n$ as a
subalgebra of $gl(n+1)$. This possibility is based on the
circumstance that the irreducible $gl(n+1)-$modules are also
$A_n-$irreducible. On the other hand, every irreducible
representation of $A_n$ in $W$ can be continued in infinitely
many ways to an irreducible representation of $gl(n+1)$ in the
same space. For this purpose it is enough to define the operator
$f_0=h_0+h_1+\l +h_n$ in $W$ where $h_0,h_1,\l,h_n$ is the
covariant basis (28) in $\tilde{\H}$. Since $f_0$ commutes with
$gl(n+1)$, $f_0$ has to be a constant in $W$, i.e.,
$$
f_0x=\Lambda_0 x \q \forall x\in W \e(55)
$$
with $\Lambda_0$ being an arbitrary number. Let $f_1,\l,f_n$ be the
$F-$ basis (34) in $\H$. Then $$
\tilde{F}=\{f_0,f_1,\l,f_n\}  \e(56)
$$
defines a basis in the Cartan subalgebra $\Ht \subset gl(n+1)$.

The eigenvalues $\La_0,\La_1,\l,\La_n$ of $\Ft$ on the highest
weight vector $x_\La \in W$ characterize $W$ as an irreducible
$gl(n+1)-$module. Let $x_{\la_1},\l,x_{\la_N}$ be a basis of
weight vectors in $W$. In view of (55) the $A_n$-weights
$\la_1,\l,\la_N$ are naturally extended to linear functional on
$\Ht$ from the requirement $\la_i(f_0)=\La_0$. Then for any
weight vector $x_\la$ we have
$$
hx_\la=\la(h)x_\la=(h,\la)x_\la \q h\in\Ht. \e(57)
$$
The numbers  $\La_0,\La_1,\l,\La_n$ are the co-ordinates of the
highest weight $\La$ in the basis
$$
\Kt=\{f^0,f^1,\l,f^n\}  \e(58)
$$
dual to $\Ft$. We call $\Kt$ {\it a canonical basis} of $gl(n+1)$ and
the co-ordinates $[\La_0,\La_1,\l,\La_n]$ - {\it canonical
co-ordinates} of the $gl(n+1)-$module $W$. The properties of the
Weyl group, which we shall often use, read more simply in the
orthogonal contravariant basis $h^0,h^1,\l,h^n$. From the equality
$$
\La=\sum_{i=0}^n\La_i f^i =\sum_{i=0}^n L_i h^i
$$
we obtain for the orthogonal co-ordinates $L_0,L_1,\l,L_n$ of the
highest weight of $W$ the following expressions
$$
\eqalign{
& L_0={1\over{n+1}}[\La_0+n\La_1+(n-1)\La_2+\l+1.\La_n] \cr
& L_1=L_0-\La_1,\cr
& L_2=L_0-\La_1-\La_2,\cr
& ...................................\cr
& L_n=L_0-\La_1-\La_2-\l-\La_n.\cr
}\e(59)
$$

Since in the $A_n-$module $W$ $\La_1,\l,\La_n$ are non-negative
integers and $\La_0$ is an arbitrary constant, it can be chosen
such that all orthogonal co-ordinates $L_0,L_1,\l,L_n$ are integers.
Moreover
$$
L_0 \ge L_1 \ge L_2 \ge \l \ge L_n. \e(60)
$$

We pass now to the main problem of this section, classification
of the Fock spaces. Unless otherwise stated, the roots and the
weights are represented by their orthogonal co-ordinates in the
contravariant orthogonal basis $h^0,h^1,\l,h^n$ in $\Ht$, i.e.,
$$
\la=(l_0,l_1,\l,l_n)\equiv \sum_{i=0}^n l_i h^i.  \e(61)
$$

\bigskip
\n
{\it Theorem 2.} The irreducible $A_n-$module is a Fock space if
and only if its highest weight is $(p,0,\l,0)$; $p$ is an
arbitrary positive integer.\footnote{*}{the case $p=0$
corresponds to the trivial one-dimensional representation.}

\smallskip
\n
{\it Proof.} As we know (theorem 1), the Fock spaces are those
and only those irreducible $A_n-$modules whose highest weight
vectors $x_\La$ are annihilated by all operators
$a_i^-a_j^+,\; i\ne j=1,\l,n$, i.e.,
$$
a_i^-a_j^+x_\La=0 \q i\ne j=1,\l,n. \e(62)
$$
Since $a_i^- x_\La=0$ and $[a_i^-,a_j^+x_]=e_{ij}$ (62) can be
replaced by the requirement
$$
e_{ij} x_\La=0 \q i\ne j=1,\l,n. \e(63)
$$
The generators $e_{ij}$ are root vectors of $A_n$ with roots
(31), i.e.,
$$
e_{ij} \; \leftrightarrow \; h^i-h^j, \q i\ne j=1,\l,n.
$$
For $i<j \; e_{ij}$ is a positive root vector and (63) holds from
the definition of $x_\La$. It remains to determine those
$A_n-$modules with weights
$$
\La=(L_0,L_1,\l,L_n) \e(64)
$$
for which the sums
$$
\La + h^j-h^i, \q i<j=1,\l,n  \e(65)
$$
are not weights.

We shall use the properties (14) and (15) of the Weyl group $S$.
According to (14) if $S_{h^i-h^j}\in S$ and
$\la=(l_0,\l,l_i,\l,l_j,\l,l_n)$ is a weight, then
$S_{h^i-h^j}.\la$ is also a weight. Using the scalar product (29)
we have
$$
S_{h^i-h^j}.\la=\la - {2(\la,h^i-h^j)\over{(h^i-h^j,h^i-h^j)}}
(h^i-h^j)=(l_0,\l,(l_j)_i,\l,(l_i)_j,\l,l_n) \e(66)
$$
where $(l_j)_i$ (resp. $(l_i)_j)$ on the r.h.s. of (66) is to
indicate that $l_j$ (resp. $l_i$) is situated on the place $i$
(resp. $j$), whereas any other $l_k$ is on the place $k$.

  Thus, the Weyl group in this case reduces to (all
possible) permutations of the orthogonal co-ordinates. For the
highest weight (64) we have
$$
\eqalign{
& S_{h^i-h^j}.\La=(L_0,\l,(L_j)_i,\l,(L_i)_j,\l,L_n)=\cr
& (L_0,\l,L_i,\l,L_j,\l,L_n)+
  (L_i-L_j)(0,\l,0,(-1)_i,0,\l,0,(1)_j,0,\l,0).  \cr
}
$$
According to (15) all vectors
$$
(L_0,\l,L_i,\l,L_j,\l,L_n)+
  k(0,\l,0,(-1)_i,0,\l,0,(1)_j,0,\l,0) \e(67)
$$
with $0\le k \le L_i-L_j$ are also weight. As we know, for
$i<j \;\; L_i\ge L_j$. Suppose $L_i>L_j.$ Then $k$ in (67) can be
equal to one and
$$
\la=\La + h^j-h^i,\q i<j
$$
is a weight. Hence the $A_n-$module $W$ is not a Fock space if in
its orthogonal signature $\La = (L_0,L_1,\l,L_n)$ there exists
$L_i>L_j$ for $0<i<j$.

It remains to consider the modules with
$$
\La = (L_0,L,\l,L), \q L_0\ge L. \e(68)
$$
Suppose for  $0<i<j$
$$
\la=\La + h^j-h^i=(L_0,L,\l,L,(L-1)_i,L,\l,L,(L+1)_j,L,\l,L)
$$
is a weight. Then
$$
\la'=(L_0,L,\l,L,(L+1)_i,L,\l,L,(L-1)_j,L,\l,L)
$$
is also a weight. This is, however, impossible since $\la'>\La$. Hence
all $A_n-$modules with signatures (68) are Fock spaces.

We could have stopped the proof here since the signatures
$$
(L_0,L,\l,L) \q {\rm and} \q (L_0-L,0, \l, 0) \e(69)
$$
describe one and the same $A_n-$module. This could have been done
if all information was carried by $A_n$, i.e., if the dynamical
variables were functions of the generators of $A_n$ only. This is
however not the case. The 4-momentum (42) $P^m \notin A_n$
although $P^m\in gl(n+1)$. Therefore physically the
representations (69) are distinguishable.

We shall determine the orthogonal co-ordinates of $\La$ from the
requirement for the energy of the vacuum to be zero. In terms of
the orthogonal basis (28) $P^m$ can be written as
$$
P^m=\sum_{i=1}^n k_i^m h_i. \e(70)
$$
Since for $\La = (L_0,L,\l,L) \q h_ix_\La=Lx_\La,\;\; i=1,\l,n$,
we require
$$
P^m |0\ra =\sum_{i=1}^n k_i^m L|0\ra=0, \q m=1,2,3. \e(71)
$$
Here $k_1^0,\l,k_n^0$ are analogue of the energy spectrum of the
one-particle states, $k_i^0>0$ (see (18)). Therefore (71) implies
$L=0$.

Later on we shall see that $h_i,\; i=1,\l,n$ is a number operator
for particles in a state $i$. This together with (71) also gives
$L=0$.

Consider the Fock space $W_p$ with $\La=(p,0,\l,0)$. Using the
definition (41) of the $a-$operators, from (62) we have
$$
a_i^-a_j^+|0\ra=p|0\ra. \e(72)
$$
We obtain the same expression as in the case of parastatistics of
order $p$ [8]. Therefore we call $p$ an order of the
$A-$statistics. We conclude that like in the parastatistics all
Fock spaces are labelled with positive integers $p$, the order of
the statistics.

The equation (72) together with the commutation relations (44) of
the $a-$operators determines completely the representation space
of the creation and annihilation operators of order $p$. The
$A-$statistics can be defined by the relations (44). The
representations of the statistics can be obtained from (72). In
this case all calculations can be done without using any Lie
algebraical properties of the $a-$operators. Clearly this point
of view is convenient for generalization to the case of infinite
and in particular to continuum number of operators. The Lie
algebraical structure however helps a lot in all calculations.
Therefore we shall continue to consider a finite number of pairs
$a_1^\pm,\l,a_n^\pm$ of $a-$operators and on a later stage we
shall let $n \rightarrow \infty$.

Let us consider some Lie-algebraical properties of the Fock spaces.
In the $A_n-$module $W$ with a highest weight
$\La =(L_0,L_1,\l,L_n)$ an arbitrary weight $\la
=(l_0,l_1,\l,l_n)$  can be represented as
$$
\la=\La + \sum_i k_i \w_i, \q k_i \in N_0,
$$
where
$$
\w_i \in \Sigma^- = \{h^i-h^j | i>j=0,1,\l,n \}.
$$
Since the sum of the first $m$ co-ordinates, $m=1,2,\l,n$ of an
arbitrary negative root is non-positive this is true also for the
vector $\sum_i k_i \w_i$ with $k_i$ non-negative integers.
Therefore for an arbitrary weight $\la$ we have
$$
l_0+l_1+\l+l_m \le L_0+L_1+\l+L_m, \q m=0,1,\l,n.
$$

From this inequality and the circumstance that the weight system
is invariant under the permutation of the orthogonal co-ordinates
we conclude that the vector $\la=(l_0,l_1,\l,l_n)$ with integer
co-ordinates is a weight if
$$
l_{i_0}+l_{i_1}+\l+l_{i_m} \le L_0+L_1+\l+L_m \e(73)
$$
where $i_0\ne i_1\ne \l,\ne i_m=0,1,\l,n;\;\; m=0,1,\l,n$.
Clearly (73) is equality for $m=n$.

\bigskip
\n
{\it Lemma 3.} All weights of the $A_n-$module of Fock $W_p$ with
order of the statistics $p$ are simple.

\smallskip
\n
{\it Proof.}
An arbitrary weight vector $x_\la \in W_p$ with a weight $\la$ is
generated from $x_\La$ with polynomials of the creation operators,
$$
x_\la = P(a_1^+,\l,a_n^+)x_\La. \e(74)
$$
Therefore the weight $\la=(l_0,l_1,\l,l_n)$ of $x_\la$ can be
represented as
$$
\la=\La + \sum_{i=1}^n k_i(-l^0+h^i), \q k_i \in N_0. \e(75)
$$
In terms of the co-ordinates the last relation reads as
$$
(l_0,l_1,\l,l_n)=(p,0,\l,0)+(-\sum_{i=1}^n k_i,k_1,k_2,\l,k_n).\e(76)
$$
Hence $k_i=l_i,\; i=1,\l,n$ and an arbitrary weight $\la$ is
represented uniquely in the form (75). In terms of the weight
vectors this gives that
$P(a_1^+,\l,a_n^+)$ in (74) is homogeneous with respect to every
creation operator $a_i^+$:
$$
P(a_1^+,\l,\alpha a_i^+,\l, a_n^+)=
\alpha^{l_i}P(a_1^+,\l,a_i^+,\l, a_n^+).
$$
Since the creation operators commute,
$$
P(a_1^+,a_2^+,\l,a_n^+)=(a_1)^{l_1}(a_2)^{l_2}\l(a_n)^{l_n}.
$$
Therefore every vector $x_\la$ with a weight $\la=(l_0,l_1,\l,l_n)$
is collinear to the vector
$$
(a_1^+)^{l_1}(a_2^+)^{l_2}\l(a_n^+)^{l_n}x_\La
$$
and the corresponding weight space is one-dimensional.

\smallskip
This lemma has no analogy in the parastatistics. For instance the
states $b_i^+b_j^+|0\ra$ and $b_j^+b_i^+|0\ra,\;i\ne j$ have one
and the same weight but in general are linearly independent.

In the following lemma we prove one important property of the
$A-$statistics.

\bigskip
\n
{\it Lemma 4.} Given $A_n-$module of Fock $W_p$ with an order of
the statistics $p$. The vector
$$
(a_1^+)^{l_1}(a_2^+)^{l_2}\l(a_n^+)^{l_n}|0\ra  \e(76')
$$
is not zero if and only if
$$
l_1+l_2+\l+l_n\le p. \e(77)
$$
In particular in the Fock space $W_p$ there can be no more than $p$
particles.

\smallskip
\n
{\it Proof.}
In the previous lemma we saw that the vector (76') has a weight
$$
\la=(p-l_1-\l-l_n,l_1,\l,l_n). \e(78)
$$
If $l_1+\l+l_n \le p$, then clearly (73) holds because
$L_0+\l+L_m=p, \; m=0,1,\l,n$. Therefore $\la$ is a weight. There
should exists at least one weight vector with weight $\la$. Since
the multiplicity of $\la$ is one, this is the vector (76') and
hence this vector is not zero.

If $l_1+\l+l_n>0$, the weight (78) does not fulfil the inequality
(73) for $m=n-1$ and
$l_{i_0}=l_1,\;l_{i_1}=l_2,\,l_{i_{n-1}}=l_n$ and the
corresponding weight vector (76') is zero.

From (76') and (78) we conclude
$$
h_i(a_1^+)^{l_1}(a_2^+)^{l_2}\l(a_n^+)^{l_n}|0\ra=
l_i(a_1^+)^{l_1}(a_2^+)^{l_2}\l (a_n^+)^{l_n}|0\ra, \q
i=1,\l,n.  \e(79)
$$
The operator $h_i$ is the number operator $N_i$ of the particles in the
state $i$. The number operator $N$ is
$$
N=N_1+N_2+\l+N_n. \e(80)
$$
We obtain
$$
N(a_1^+)^{l_1}(a_2^+)^{l_2}\l(a_n^+)^{l_n}|0\ra=
(l_1+\l +l_n)(a_1^+)^{l_1}(a_2^+)^{l_2}\l(a_n^+)^{l_n}|0\ra.\e(81)
$$

\vfill\eject

\n
{\bf 4. Matrix elements of the creation and annihilation operators }

\bigskip
The numbers $l_1,\l,l_n$ together with the order of the
$A-$statistics $p$ determine uniquely the state (76'). We
introduce the notation
$$
|p;l_1,l_2,\l,l_n\ra=(a_1^+)^{l_1}(a_2^+)^{l_2}\l(a_n^+)^{l_n}|0\ra.\e(82)
$$
The set of all vectors (82) constitute a basis of weight vectors
in the Fock space $W_p$. The correspondence between the weight
vectors and the weights written in their orthogonal co-ordinates
reads as
$$
|p;l_1,l_2,\l,l_n\ra \q \leftrightarrow \q
(p-\sum_{i=1}^n l_i,l_1,l_2,\l,l_n).  \e(83)
$$
One has to remember that the notation $|p;l_1,l_2,\l,l_n\ra$ is
defined for $l_1+\l +l_n \le p.$

We now proceed to calculate the matrix elements on $n$ pairs of
creation and annihilation operators $a_1^\pm,\l,a_n^\pm$ in the
$A_n-$module of Fock $W_p$ with order of the statistics $p$.

We can write immediately
$$
\eqalign{
& h_0 |p;l_1,l_2,\l,l_n\ra=(p-\sum_{i=1}^n l_i)|p;l_1,l_2,\l,l_n\ra,\cr
& h_i |p;l_1,l_2,\l,l_n\ra=l_i|p;l_il_1,l_2,\l,l_n\ra,
\q i=1,\l,n.\cr
}\e(84)
$$
These equations follow from the observation that the orthogonal
co-ordinates of the weight (83) are eigenvalues of the operators
(28) on the weight vector (82). Since
$$
[a_i^-,a_i^+]=h_0-h_i
$$
we have
$$
[a_i^-,a_i^+]|p;l_1,l_2,\l,l_n\ra =
(p-L-l_i)|p;l_1,l_2,\l,l_n\ra, \e(85)
$$
where $L=l_1+l_2+\l+l_n$.

First we calculate the matrix elements of $a_1^-$.
$$
\eqalign{
& a_1^-|p;l_1,l_2,\l,l_n\ra=
[a_1^-,(a_1^+)^{l_1}(a_2^+)^{l_2}\l(a_n^+)^{l_n}]|0\ra\cr
& =[a_1^-,(a_1^+)^{l_1}](a_2^+)^{l_2}\l(a_n^+)^{l_n}|0\ra+
(a_1^+)^{l_1}a_1^-(a_2^+)^{l_2}\l(a_n^+)^{l_n}|0\ra.\cr
}\e(86)
$$
The second term in the last equality vanishes. Indeed the vector
$$
a_1^-(a_2^+)^{l_2}\l(a_n^+)^{l_n}|0\ra
$$
would have had a weight
$$
(p-\sum_{i=2}^n+1,-1,l_2,l_3,\l,l_n)
$$
which is impossible since $l_0+l_2+l_3+\l+l_n=p+1>p$.

Using (84), for the first term we obtain
$$
\eqalign{
& \sum_{i=0}^{l_1-1} (a_1^+)^i [a_i^-,a_1^+](a_1^+)^{l_1-i-1}
  (a_2^+)^{l_2}\l(a_n^+)^{l_n}|0\ra= \cr
& \sum_{i=0}^{l_1-1}(p-L-l_1+2i+2)(a_1^+)^{l_1-1}
  (a_2^+)^{l_2}\l(a_n^+)^{l_n}|0\ra . \cr
}
$$
This gives
$$
a_1^-|p;l_1,l_2,\l,l_n\ra=
l_1(p-\sum_{i=1}^n l_i+1)|p;l_1-1,l_2,\l,l_n\ra.
$$
The generalization for $a_i^-$ is evident:
$$
a_i^-|p;l_1,\l l_{i-1},l_i,l_{i+1},\l,l_n\ra=
l_i(p-\sum_{k=1}^n l_k+1)|p;l_1\l,l_{i-1},l_i-1,l_{i+1}\l,l_n\ra.
\e(87)
$$
Moreover
$$
a_i^+|p;l_1,\l l_{i-1},l_i,l_{i+1},\l,l_n\ra=
|p;l_1\l,l_{i-1},l_i+1,l_{i+1}\l,l_n\ra.
\e(88)
$$
The metric in $W_p$ is defined in a complete analogy with the
scalar product in the Fock space of Bose (or Fermi) operators.

Postulate
$$
\eqalign{
& a) \q \langle 0|0\ra=1, \cr
& b) \q \langle 0|a_i^+ =0, \q i=1,\l,n,\cr
& c) \q ((a_1^+)^{m_1}(a_2^+)^{m_2}\l(a_n^+)^{m_n}|0\ra,
     (a_1^+)^{l_1}(a_2^+)^{l_2}\l(a_n^+)^{l_n}|0\ra)=\cr
& \q\q	 =\langle 0|(a_1^-)^{m_1}(a_2^-)^{m_2}\l(a_n^-)^{m_n}
     (a_1^+)^{l_1}(a_2^+)^{l_2}\l(a_n^+)^{l_n}|0\ra).\cr
}\e(89)
$$

The vectors $|p;l_1,\l,l_n\ra$ constitute an orthogonal basis in
$W_p$. To show this, suppose that in (89) some $m_i \ne l_i$ and let
$m_i>l_i$. Then the vector
$$
(a_i^-)^{m_i}(a_1^+)^{l_1}\l(a_i^+)^{l_i}\l(a_n^+)^{l_n}|0\ra)=0
$$
since otherwise there has to exists a weight
$$
(p-\sum_{j=1}^n l_j+m_i,l_1,\l,l_{i-1},-(m_i-l_i),l_{i+1},\l,l_n )
$$
which is impossible. For $m_i<l_i$ the same result can be
obtained from the hermitian conjugate of (89). If
$m_i=l_i,\;i=1,\l,n$ we obtain
$$
(|p;l_1,\l,l_n\ra,|p;l_1,\l,l_n\ra)=
{p!\over (p-L)!}\prod_{i=1}^n l_i!, \e(90)
$$
where $L=l_1+\l+l_n$.

As an orthogonal basis in $W_p$ one can accept the vectors
$$
|p;l_1,\l,l_n)=\sqrt{(p-L)!\over p!}
{(a_1^+)^{l_1}\l(a_n^+)^{l_n}\over{\sqrt{l_1!l_2!\l l_n!}}}|0\ra.\e(91)
$$
In this basis we have for the matrix elements
$$
\eqalignno{
& a_i^+|p;l_1,\l l_{i-1},l_i,l_{i+1},\l,l_n)=
  \sqrt{(l_i+1)(p-\sum_{j=1}^n l_j  )}
|p;l_1\l,l_{i-1},l_i+1,l_{i+1}\l,l_n). & (92)\cr
& a_i^-|p;l_1,\l l_{i-1},l_i,l_{i+1},\l,l_n)=
  \sqrt{l_i(p-\sum_{j=1}^n l_j +1  )}
|p;l_1\l,l_{i-1},l_i-1,l_{i+1}\l,l_n). & (93)\cr
}
$$

The matrix elements of the $a-$operators do not depend on $n$.
Therefore the result can be extended in an evident way to the
case of infinite number of operators.

Finally, we point out one interesting property of the $A-$statistics.
Introduce the operators
$$
A_i^\pm = {a_i^\pm\over \sqrt{p}}, \q i=1,\l,n  \e(94)
$$
and consider the matrix elements of these operators on states with
number of particles much less then $p$,
$$
l_1+l_2+\l+l_n << p. \e(95)
$$
From (92-93) we obtain
$$
\eqalign{
& a_i^-|p;l_1,\l l_{i-1},l_i,l_{i+1},\l,l_n) \simeq
  \sqrt{l_i}\; |p;l_1\l,l_{i-1},l_i-1,l_{i+1}\l,l_n), \cr
& \cr
& a_i^+|p;l_1,\l l_{i-1},l_i,l_{i+1},\l,l_n) \simeq
  \sqrt{l_i+1}\;|p;l_1\l,l_{i-1},l_i+1,l_{i+1}\l,l_n). \cr
} \e(96)
$$
In a first approximation
$$
\eqalign{
& [A_i^+,A_j^+]=[A_i^-,A_j^-],\q {\rm exact \;  commutators}, \cr
& [A_i^-,A_j^+]=\delta_{ij}, \q {\rm if}\; l_1+l_2+\l+l_n << p.
}\e(97)
$$
Moreover if (95) holds, then
$$
|p;l_1,\l,l_n)=
{(A_1^+)^{l_1}\l(A_n^+)^{l_n}\over{\sqrt{l_1!l_2!\l l_n!}}}|0\ra.\e(98)
$$
We see that if the $A-$statistics allows a large number of
particles $p$, then the commutation relations of the operators
$A_i^\pm$ on states with $ l_1+l_2+\l+l_n << p$ coincide in a first
approximation with the Bose creation and annihilation operators.
In the limit $p \rightarrow \infty$ the operators $A_i^\pm$
reduce to Bose operators.

This property has also an interesting Lie-algebraical
consequence. It shows that the limits of certain representations
(the Fock representations) of the simple algebra $A_n$ leads to a
representation of the solvable Lie algebra of Bose operators.

We have considered the statistics corresponding to the algebra of
the unimodular group. In a similar way one can introduce a
concept of $C-$ and $D-$statistics [9] or of statistics that
correspond to other semisimple Lie algebras [5].

\vskip 24pt
\noindent
{\bf References}

\vskip 12pt
\settabs \+  [11] & I. Patera, T. D. Palev, Theoretical
   interpretation of the experiments on the elastic \cr

\+ [1] & Green H S 1953 {\it Phys. Rev.} {\bf 90} 270 \cr

\+ [2] & Kamefuchi S and Takahashi Y 1960 {\it Nucl. Phys.}
         {\bf 36} 177\cr

\+ [3] & Ryan C and Sudarshan E C G 1963 {\it Nucl. Phys.}
         {\bf 63} 207 \cr

\+ [4] & Palev T D 1975 {\it Ann. Inst. H. Poincar\'e}
         {\bf 13} 49 \cr

\+ [5] & Palev T D 1976 {\it Preprint JINR} E2-10258, Dubna\cr

\+ [6] & Bogoljubov N N, Shirkov D V {\it Introduction to the Theory
         of Quantized Fields},\cr
\+     & Moscow 1957 (English ed. Interscience Publishers, Inc.,
         New York, 1959)\cr

\+ [7] & Jacobson N {\it Lie Algebras} (Interscience Publishers, Inc.,
         New York, 1962)\cr

\+ [8] & Greenberg O W, Messiah A M 1965 {\it Phys. Rev.}
         {\bf 138} 1155 \cr

\+ [9] & Palev T D, {\it Thesis}, Institute for Nuclear Research and
         Nuclear Energy, Sofia (1976)\cr

\end